\documentclass{article}

\usepackage{arxiv}
\usepackage[utf8]{inputenc} % allow utf-8 input
\usepackage[T1]{fontenc}    % use 8-bit T1 fonts
\usepackage{hyperref}       % hyperlinks
\usepackage{url}            % simple URL typesetting
\usepackage{booktabs}       % professional-quality tables
\usepackage{amsfonts}       % blackboard math symbols
\usepackage{nicefrac}       % compact symbols for 1/2, etc.
\usepackage{microtype}      % microtypography
\usepackage{lipsum}
\usepackage{amsmath}
\usepackage{graphicx}

\usepackage{natbib}
\setcitestyle{authoryear, open={(},close={)}}
\def\d{{\, \rm d}}

\makeatletter
\def\ps@myheadings{%
    \let\@oddfoot\@empty\let\@evenfoot\@empty
    \def\@evenhead{\thepage\hfil\slshape\leftmark}%
    \def\@oddhead{{\slshape\rightmark}\hfil\thepage}%
    \let\@mkboth\@gobbletwo
    \let\sectionmark\@gobble
    \let\subsectionmark\@gobble
    }
  \if@titlepage
  \renewcommand\maketitle{\begin{titlepage}%
  \let\footnotesize\small
  \let\footnoterule\relax
  \let \footnote \thanks
  \null\vfil
  \vskip 60\p@
  \begin{center}%
    {\LARGE \@title \par}%
    \vskip 3em%
    {\large
     \lineskip .75em%
      \begin{tabular}[t]{c}%
        \@author
      \end{tabular}\par}%
      \vskip 1.5em%
    {\large \@date \par}%       % Set date in \large size.
  \end{center}\par
  \@thanks
  \vfil\null
  \end{titlepage}%
  \setcounter{footnote}{0}%
}
\else
\renewcommand\maketitle{\par
  \begingroup
    \renewcommand\thefootnote{\@fnsymbol\c@footnote}%
    \def\@makefnmark{\rlap{\@textsuperscript{\normalfont\@thefnmark}}}%
    \long\def\@makefntext##1{\parindent 1em\noindent
            \hb@xt@1.8em{%
                \hss\@textsuperscript{\normalfont\@thefnmark}}##1}%
    \if@twocolumn
      \ifnum \col@number=\@ne
        \@maketitle
      \else
        \twocolumn[\@maketitle]%
      \fi
    \else
      \newpage
      \global\@topnum\z@   % Prevents figures from going at top of page.
      \@maketitle
    \fi
    \thispagestyle{plain}\@thanks
  \endgroup
  \setcounter{footnote}{0}%
}
\makeatother

\title{A Bayesian Machine Learning Algorithm for Predicting ENSO Using Short Observational Time Series}

\author{
  Nan Chen \\
  Department of Mathematics\\ University of Wisconsin-Madison\\ Madison, WI 53706, USA\\
   \And
  Faheem Gilani \\
  Department of Mathematics\\ The Pennsylvania State University\\ University Park, PA 16802, USA\\
  \And
  John Harlim \\
  Department of Mathematics\\Department of Meteorology and Atmospheric Science\\ Institute for Computational and Data Sciences\\The Pennsylvania State University\\ University Park, PA 16802, USA\\}

\begin{document}
\maketitle

\begin{abstract}% word limit: 150
A simple and efficient Bayesian machine learning (BML) training  algorithm, which exploits only a 20-year short observational time series and an approximate prior model, is developed to predict the Ni\~no 3 sea surface temperature (SST) index. The BML forecast significantly outperforms model-based ensemble predictions and standard machine learning forecasts. Even with a simple feedforward neural network, the BML forecast is skillful for 9.5 months. Remarkably, the BML forecast overcomes the spring predictability barrier to a large extent: the forecast starting from spring remains skillful for nearly 10 months. The BML algorithm can also effectively utilize multiscale features: the BML forecast of SST using SST, thermocline, and windburst improves on the BML forecast using just SST by at least $2$ months. Finally, the BML algorithm also reduces the forecast uncertainty of neural networks and is robust to input perturbations.

\end{abstract}
\noindent\textbf{Key words:} Bayesian machine learning; Short observations; Spring predictability barrier; Forecast uncertainty

\section*{Key Points}
\begin{itemize}
\item A new Bayesian machine learning (BML) framework is developed to accommodate  the shortage of observations when training neural networks.
\item The new BML forecast significantly  outperforms  model-based ensemble predictions and standard machine learning forecasts.
\item The new BML algorithm reduces forecast uncertainty and overcomes the spring predictability barrier to a large extent.
\end{itemize}
\section*{Plain Language Summary}% word limit: 200
One major challenge in applying machine learning algorithms for predicting the El Ni\~no Southern Oscillation (ENSO) is the shortage of observational training data. In this article, a simple and efficient Bayesian machine learning (BML) training   algorithm is developed, which exploits only a 20-year observational time series for training a neural network. In this new BML algorithm, a long simulation from an approximate parametric model is used as the prior information while the short observational data plays the role of the likelihood which corrects the intrinsic model error in the prior data during the training process. The BML algorithm is applied to predict the Ni\~no 3 sea surface temperature (SST) index. Forecast from the BML algorithm outperforms standard machine learning forecasts and model-based ensemble predictions. The BML algorithm also allows a multiscale input consisting of both the SST and the wind bursts that greatly facilitate the forecast of the Ni\~no 3 index.  Remarkably, the BML forecast overcomes the spring predictability barrier to a large extent. Moreover, the BML algorithm reduces the forecast uncertainty and is robust to the input perturbations.

%% ------------------------------------------------------------------------ %%
%
%  TEXT
%
%% ------------------------------------------------------------------------ %%

%%% Suggested section heads:
% \section{Introduction}
%
% The main text should start with an introduction. Except for short
% manuscripts (such as comments and replies), the text should be divided
% into sections, each with its own heading.

% Headings should be sentence fragments and do not begin with a
% lowercase letter or number. Examples of good headings are:

% \section{Materials and Methods}
% Here is text on Materials and Methods.
%
% \subsection{A descriptive heading about methods}
% More about Methods.
%
% \section{Data} (Or section title might be a descriptive heading about data)
%
% \section{Results} (Or section title might be a descriptive heading about the
% results)
%
% \section{Conclusions}

\section{Introduction}
As the most prominent interannual climate variability, the El Ni\~no Southern Oscillation (ENSO) manifests as a basin-scale air-sea interaction phenomenon characterized by sea surface temperature (SST) anomalies in the equatorial central to eastern Pacific \cite{clarke2008introduction, zebiak1987model, rasmusson1982variations}. It has a strong impact on climate, ecosystems, and economies around the world through global circulation \cite{ropelewski1987global,ashok2009nino}. Classically, ENSO is regarded as a  cyclic phenomenon \cite{wyrtki1975nino,jin1997equatorial}, in which the positive and negative phases are known as El Ni\~no and La Ni\~na, respectively.

The traditional ensemble forecast using physics-based models has been widely used for predicting the ENSO \cite{moore1998skill, tang2018progress, kirtman2009multimodel}. A hierarchy of models ranging from the general circulation models (GCMs) to many intermediate and low-order models are employed for forecasting the refined and large-scale ENSO features, respectively. However, model error, which leads to large predictive uncertainty, is ubiquitous in these parametric models and often results in ineffective forecasts. The model error often comes from the incomplete understanding of nature and/or the inadequate spatiotemporal resolutions in these models \cite{palmer2001nonlinear, kalnay2003atmospheric, majda2018model}. More recently, machine learning techniques have become prevalent in forecasting ENSO and many other climate phenomena \cite{ding2018skillful, ham2019deep, lecun2015deep, wang2020extended}. These machine learning approaches exploit sophisticated neural networks or other nonparametric methods to recover the complex dynamics in nature. Given sufficient training data, these approaches can achieve state-of-the-art numerical performance, beating traditional physics-based models.

However, one major challenge in applying machine learning algorithms for predicting ENSO is the shortage of observational training data. In fact, only three extreme El Ni\~no events and a few moderate ones were observed during the satellite era (i.e., from 1980 to the present). To augment the training data set, commonly used strategies include concatenating the satellite observations with either the proxy-based reconstructed data \cite{rayner2003global, barrett2018reconstructing, gergis2009history, mcgregor2010unified, emile2013estimating} or with time series generated from certain parametric models (such as a GCM) \cite{ham2019deep}. Yet, the augmented data from both sources often contain a range of uncertainties and inaccuracies when compared with the high-resolution satellite observations for characterizing the ENSO features. Therefore, developing a new machine learning training algorithm that systematically reduces the forecasting errors and uncertainties in the augmented training data set becomes essential for extending  ENSO forecasting skills.

In this article, we develop a new Bayesian machine learning (BML) training  algorithm that utilizes only short observational time series for an effective prediction of the ENSO. The focus here is on predicting the Ni\~no 3 SST index, which is a commonly used ENSO index for characterizing the large-scale features of the eastern Pacific El Ni\~no. A simple feedforward neural network \cite{fine2006feedforward} is adopted as the prediction model. In this BML algorithm, a long simulation from a parametric model is used as the prior information for training the neural network while the short observational data plays the role of the likelihood which corrects the intrinsic model error in the prior data \cite{bernardo2009bayesian, box2011bayesian} during the training process. The neural network model trained using the BML framework outperforms the same neural network model trained with the standard procedure and model-based ensemble predictions. Note that the new BML algorithm is adaptable to any neural network architecture and any geophysical system with limited observations \cite{watson2017impact}, provided that a reasonable approximate parametric model is in hand.

The rest of the article is organized as follows. The observational data sets are described in Section \ref{Sec:DataSets}. A simple three-dimensional parametric model that is utilized to generate the long time series as the prior information for training the neural network is introduced in Section \ref{Sec:3DModel}. The new BML training  algorithm is developed in Section \ref{Sec:Algorithm}. The forecast results are shown in Section \ref{Sec:Results}. The article is concluded in Section \ref{Sec:Conclusion}.

\section{The Observational Data Sets}\label{Sec:DataSets}
In this study, we use reanalysis data from satellite observations. The SST data is from the Optimum Interpolation Sea Surface Temperature (OISST) reanalysis \cite{reynolds2007daily} while the zonal winds are at 850hPa from the National Centers for Environmental Prediction/National Center for Atmospheric Research (NCEP/NCAR) reanalysis \cite{kalnay1996ncep}. The thermocline depth is computed from the potential temperature as the depth of the 20$^o$C isotherm using the NCEP Global Ocean Data Assimilation System (GODAS) reanalysis data \cite{behringer1998improved}. The temporal resolution of all the data is daily and they cover the period from 1982/01/01 to 2020/02/29. The spatial resolutions are 0.25$^o$, 1$^o$, and 2.5$^o$, respectively, for the SST, the thermocline depth, and the zonal winds.
All datasets are averaged meridionally within 5N-5S in the tropical Pacific (120E-80W), which is followed by removing the climatology mean and the seasonal cycle. %A 90-day running average is then applied for the SST and the thermocline depth data.

The following three indices are derived from the above data sets: the averaged SST in the Ni\~no 3 region (150W-90W; which is essentially the eastern Pacific) $T_E$, the averaged thermocline depth in the western Pacific region (120E-160W) $H_W$, and the averaged wind bursts over the western Pacific region $\tau$.

The Ni\~no 3 SST index $T_E$ characterizes the eastern Pacific El Ni\~nos and rules out most of the central Pacific events \cite{di2010central, yeh2009nino}. The main reason for choosing such a simple index as the prediction target is that the prior information of $T_E$ is naturally obtained from the recharge-discharge paradigm \cite{jin1997equatorial}, which facilitates
the understanding of the BML algorithm. The BML algorithm can be easily applied to predict the Ni\~no 3.4 index and the spatiotemporal evolutions of the ENSO.

\section{A Simple Non-Gaussian Parametric Model for the Prior Information}\label{Sec:3DModel}
The BML algorithm trains the neural network model in a Bayesian framework, using a long time series generated from a (prior) parametric model. To this end, the parametric model and the associated time series are called the ``prior parametric model'' and the ``prior data'', respectively. Note that this prior parametric model does not need to be perfect, as it is often the case in practice. Yet, it has to be reasonable in the following sense. While the prior model alone may not produce an effective predictive skill, it will provide a set of training data that reflects some qualitative features of the input variables and climatological statistics, which in turn, produces  skillful forecasts when the neural-network model is trained using the proposed BML algorithm, as we shall see.
%A reasonable prior model that approximates nature is sufficient for the BML algorithm to provide effective forecast results.
%\st{As will be shown in Section} \ref{Sec:Results_ModelError}\st{, the BML forecasts are quite robust to the variation of the prior model, which facilitates a wide application of the BML algorithm in practice.}

We utilize the following three-dimensional (3D) stochastic differential equations (SDE's) as the prior model:
\begin{subequations}\label{priormodel}
\begin{align}
    \frac{\d T_E}{\d t}&=(-d_TT_E+\omega H_W+\alpha_T\tau)+\sigma_T\dot{W}_T,\label{priormodel_T}  \\
    \frac{\d H_W}{\d t}&=(-d_HH_W-\omega T_E+\alpha_H\tau)+\sigma_H \dot{W}_H, \label{priormodel_H}\\
    \frac{\d\tau}{\d t}&=(-d_\tau \tau) +\sigma_\tau(T_E)\dot{W}_\tau,  \label{priormodel_tau}
\end{align}
\end{subequations}
where, as was defined in Section \ref{Sec:DataSets}, $T_E$, $H_W$ and $\tau$ represent the averaged SST in the eastern Pacific, the averaged thermocline depth in the western Pacific, and the averaged wind bursts in the western Pacific, respectively. The constants $d_T$, $d_H$ and $d_\tau$ are the damping coefficients while the constant $\omega$ characterizes the oscillation frequency. The constants $\alpha_T$ and $\alpha_H$ are the coefficients that couple the wind bursts to the interannual variables. The two noise coefficients $\sigma_T$ and $\sigma_H$ are constants while the remaining noise coefficient $\tau(T_E)$ is a function of $T_E$. The terms $\dot{W}_T$, $\dot{W}_H$ and $\dot{W}_\tau$ are independent white noise. The parametric model in \eqref{priormodel} can be regarded as the recharge-discharge model \cite{jin1997equatorial} augmented by a random wind burst model.
Note that the increased SST anomaly enhances the convective activity, which results in more active wind bursts \cite{tziperman2007quantifying, hendon2007seasonal, puy2016modulation}. Therefore, it is essential to adopt a state-dependent (i.e., multiplicative) noise $\sigma_\tau$ in \eqref{priormodel_tau}, which assumes that the wind burst is positively correlated with the SST anomalies. Since $T_E$ is the only SST variable in \eqref{priormodel}, it is used as an approximation to the basin-averaged SST that influences the wind burst amplitudes.
The SDE \eqref{priormodel_tau} can generate both the westerly wind bursts (WWBs) and the easterly wind bursts (EWBs), corresponding to $\tau$ with positive and negative values, respectively. It is also this state-dependent noise that allows the coupled model \eqref{priormodel} to generate non-Gaussian features of the observed ENSO. One time unit in the model stands for one month. The units of $T_E$ and $\tau$ are $^o$C and m/s, respectively. On the other hand, one unit of $H_W$ in the model corresponds to $15$m, which allows $H_W$ and $T_E$ to have similar amplitudes in the model simulation.
The parameter values are determined by minimizing the errors between the prior model's and the observed time series' probability density functions (PDFs) and the autocorrelation functions (ACFs). The estimated parameters, which we will refer to as the \textbf{reference} parameters in this paper, are given as follows,
\begin{equation}\label{ParameterValues}
\begin{gathered}
  d_T=d_H=1.5,\qquad d_\tau=4,\qquad \omega=-1.5,\qquad \alpha_T=1,\qquad \alpha_H=-0.4,\\
  \sigma_T=\sigma_H=0.8,\qquad\mbox{and}\qquad \sigma_\tau(T_E)=4.5\tanh(T_E+1)+4.
\end{gathered}
\end{equation}

As will be shown in Section \ref{Sec:Results_ModelError}, the BML forecasts are quite robust to perturbations of these parameters, which allow for a wide range of application of the BML algorithm in practice.

Figure \ref{ENSO_Comparison} in the \emph{Supporting Information} shows that the prior model \eqref{priormodel} can reproduce the qualitative features of the Ni\~no 3 SST index and the associated non-Gaussian statistics. Yet, as we shall see, the model alone cannot produce effective forecasts or more complex features.  Consequently, neural networks trained using the prior data alone will not exhibit the most effective prediction skill (See Section \ref{Sec:Results_ComparisonML}). This further motivates the development of a BML framework that combines the observational time series with the model simulation to reduce the model error and improve the forecast skill of the Ni\~no 3 index.

\section{A Bayesian machine learning (BML) algorithm}\label{Sec:Algorithm}
We now discuss a Bayesian machine learning (BML)  algorithm which takes into account both the prior data described in \ref{Sec:3DModel} with the observation data described in \ref{Sec:DataSets}. Let $\mathbf{v}(t)$ be the prior data, which is a long time series from the prior parametric model \eqref{priormodel}, and $\mathbf{u}(t)$ be the short observational time series. Here $\mathbf{v}(t)$ and $\mathbf{u}(t)$ can both be one-dimensional, representing the time series of $T_E$, or they can be multi-dimensional time series containing any subset of the three variables $T_E, H_W$, and $\tau$.

Denote by $\boldsymbol{\theta}$ the parameters in the neural network and $\boldsymbol\theta_k$ the estimated parameters after the $k$-th iteration in the stochastic gradient descent (SGD) method \cite{bottou2010large}. Next, denote by $\mathbf{x}^D_i$ and $\mathbf{y}^D_i$ for $i=1,\ldots,N_D$ the input and output data, constructed from the prior time series $\mathbf{v}(t)$,
where each $\mathbf{x}^D_i$ is a delay embedded time series of $\mathbf{v}(t)$ with an appropriate delay embedding time and each $\mathbf{x}^D_i$ is a corresponding forecast value. Denote by $\mathbf{x}^O_i$ and $\mathbf{y}^O_i$ for $i=1,\ldots,N_O$ the input and output data constructed from $\mathbf{u}(t)$ in the same way that $\mathbf{x}^D_i$ and $\mathbf{y}^D_i$ are constructed from $\mathbf{v}(t)$. Since the observational time series is much shorter than the prior time series, we have $N_O\ll N_D$. Note that the input and output data will be further specified in Section \ref{Sec:Setup}.

Define two loss functions  $L^D$ and $L^O$ in training the neural network (NN). The first loss function is exploited to provide a potential update to the parameters $\boldsymbol{\theta}$,
\begin{subequations}\label{Training_NN_D}
\begin{align}
  L^D &= \sum_i\| \mathbf{y}^D_i - \mbox{NN}(\mathbf{x}^D_i; \boldsymbol{\theta})\|^2\label{Training_NN_D_a}\\
   \boldsymbol\theta_{k+1}^D &= \boldsymbol\theta_{k}^D-\eta^D_k\left.\frac{\partial L^D}{\partial\boldsymbol\theta}\right|_{\boldsymbol\theta=\boldsymbol\theta_{k}^D},\label{Training_NN_D_b}
\end{align}
\end{subequations}
where $\|\cdot\|$ in Eq.~\eqref{Training_NN_D_a} is a given metric for computing the loss function, such as the mean-squared error. We should point out one can always define a computationally reasonable loss function with a metric that is adequate for quantifying the error of the point estimator. The loss function in \eqref{Training_NN_D_a} corresponds to the standard empirical least-squares regression in supervised learning.

The equation \eqref{Training_NN_D_b} is the SGD for updating the parameters in the neural network, where $\eta^D_k$ is the standard learning rate. Note that the observational information is not involved in \eqref{Training_NN_D}; it only appears in the second loss function,
\begin{equation}\label{Training_NN_O}
  L^O = \sum_i\| \mathbf{y}^O_i - \mbox{NN}(\mathbf{x}^O_i; \boldsymbol{\theta})\|^2,
\end{equation}
which is used to validate the proposed parameters formulated in (\ref{Training_NN_D}b). The BML training algorithm contains two steps in each iteration cycle for updating $\boldsymbol\theta$. Denote by $\boldsymbol{\theta}_k^*$  the current parameter estimate. \\
{\bf Step 1: Proposal.} Generate a proposal $\boldsymbol{\theta}^D_{k+1}$ as a potential update by utilizing (\ref{Training_NN_D}b) with $\boldsymbol{\theta}^D_k=\boldsymbol{\theta}^*_k$. Note that only the prior information is used in this step.
\\
{\bf Step 2: Validation.} Evaluate the likelihood function \eqref{Training_NN_O} on the proposal $\boldsymbol\theta_{k+1}^D$ obtained from Step~1.
Then the resulting value of the loss function $L^O(\boldsymbol\theta_{k+1}^D)$ is compared with $L^O(\boldsymbol\theta_{k}^*)$. If the relationship $L^O(\boldsymbol\theta_{k+1}^D) \geq L^O(\boldsymbol\theta_{k}^*)$, we reject the proposal and repeat Step 1 with $\boldsymbol{\theta}^*_{k+1} = \boldsymbol{\theta}^*_{k}$. Otherwise, accept the proposal and let $\boldsymbol{\theta}^*_{k+1} = \boldsymbol{\theta}^D_{k+1}$ and repeat Step 1. If the proposed parameters are rejected in the last $S$ consecutive cycles, which means the proposed parameters fail to improve the loss function \eqref{Training_NN_O} (i.e., the likelihood based on the observational data), then the training process will be terminated.

We call this training procedure the Bayesian Machine Learning (BML) algorithm since the two steps above are very similar to those of the Bayesian Markov chain Monte Carlo algorithm. We should point out that the resulting predictor is a point estimator (one neural-network model). Once the model is trained via BML, the prediction of the resulting NN model is no different than the standard ML testing procedure. It is worthwhile to point out that the proposed training algorithm is not similar to the well-known Bayesian Neural Networks (BNN) \cite{mackay1995probable,lampinen2001bayesian} which has widely been used, such as in  language modelling and sentiment analysis \cite{gal2016theoretically, mukhoti2018evaluating}. The BNN aims to construct a posterior distribution of hyperparameters $\theta$, from which one can employ an ensemble of NN models with $\theta$ sampled from the estimated posterior distribution and subsequently uses the ensemble average as a point estimator or ensemble variance to quantify uncertainties of the point estimator. Technically, BNN requires a specification of the prior distribution of the hyperparameter $\theta$ and employs the optimization on the observed data (so it is not designed to overcome a shortage of observation data).

The BML algorithm is especially useful when only short observations and an approximate parametric model that contains model error are available. Note that the loss function in \eqref{Training_NN_O} is not directly used with the SGD to update $\boldsymbol{\theta}$ as would be the case when using the observation datasets to train a neural network. This is because the limited observational data results in large variance and, due to the bias-variance tradeoff \cite{geman1992neural}, larger expected error, thereby degrading forecast skill. On the other hand, although training with a sufficient amount of prior data results in lower variance and a robust training algorithm, the model error in the prior data may result in biased estimation of $\boldsymbol{\theta}$. The BLM algorithm attempts to de-bias this estimate by accepting/rejecting the proposed network parameters based on the value of the loss function on the observation data.

As a remark, the second step of the above algorithm, namely \eqref{Training_NN_O}, can also be understood as a special validation procedure \cite{mello2018machine, el2018machine} in the neural network training process. Since it utilizes only observational data, we call it the ``data-driven validation''. As will be seen in Section \ref{Sec:Results_ComparisonML}, the proposed BLM algorithm outperforms standard neural network training and validation procedures in the case of ENSO forecasting.

\section{Results}\label{Sec:Results}
\subsection{Setup}\label{Sec:Setup}
We employ a feedforward net with three hidden layers. The first two hidden layers have $32$ $\tanh$ units each and the final hidden layer has $128$ linear units. The output layer consists of $50$ linear units. The network uses SGD to minimize the root-mean-square error (RMSE) with batch sizes of $128$. In the following experiments, different input data are used, which may contain only a single variable $T_E$, two variables $(T_E, H_W)$ or three variables $(T_E, H_W, \tau)$. Each input, corresponding to an $\mathbf{x}^D_i$ in Section \ref{Sec:Algorithm}, takes into account a delay embedded time series. Here, the delay embedding time for $T_E$ and $H_W$ is the past $8$ weeks, and that for $\tau$ is the past $8$ days. Weekly data is used for $T_E$ and $H_W$ while daily data is used for $\tau$. The output contains only $T_E$.
The $i$-th output $\mathbf{y}^D_i$ represents the forecast value of $T_E$ at the lead time of the $i$-th week, where $i=1,\ldots,50$. As we stated before, $\{\mathbf{x}^O_i,\mathbf{y}^O_i\}$ will be constructed in the same way as $\{\mathbf{x}^D_i,\mathbf{y}^D_i\}$, except that the former consists of only the observed data, $\mathbf{u}(t)$. The parameter $S$ in the BML is set to be $S=20$. This parameter was determined by a quick test on validation data but the authors found that the $S=15$ and $S=25$ yield similar results so the performance of the BML algorithm using the suggested network is not too sensitive to the choice of $S=20$.

In each experiment, the training data is generated by integrating the prior model \eqref{priormodel} forward in time until there are approximately $30,000$ weekly samples of $T_E$ and $H_W$ for training (with daily windburst values chosen appropriately to match current time values of $T_E$ and $H_W$ during training). This is done by integrating forward on a daily scale and keeping every seventh observation. The efficacy of the learning algorithm is checked using a two-fold test by splitting the observation data into two time periods: first using the 1983-1999 data as training and the 2001-2020 data as testing and then vice-versa. The data in the year 2000 is excluded to prevent the overlap between the training and testing periods. The overall prediction skill is measured by evaluating the normalized RMSE (NRMSE) and pattern correlation (Corr) between the truth and the predicted time series in the entire period (1983--2020 except the year 2000). The Corr and NRMSE are defined as:
\begin{equation}\label{SkillScores}
\begin{split}
  \mbox{Corr} &= \frac{\sum_{i=1}^n(u^f_i-\bar{u}^f)(u^o_i-\bar{u}^o)}{\sqrt{\sum_{i=1}^n(u^f_i-\bar{u}^f)^2}\sqrt{\sum_{i=1}^n(u^o_i-\bar{u}^o)^2}},\\
  \mbox{NRMSE} &= 1-\frac{\mbox{RMSE}}{\mbox{std}(u^o)},\qquad\mbox{where~RMSE} =  \sqrt{\frac{\sum_{i=1}^n(u^f_i-u^o_i)^2}{n}},
\end{split}
\end{equation}
where $u^f_i$ and $u^o_i$ are the forecast and the truth, respectively, at time $t=t_i$. The time averages of the forecast and the true time series are denoted by $\bar{u}^f$ and $\bar{u}^o$ while std$(u)$ is the standard deviation of the truth.
The NRMSE starts from NRMSE $=1$ and loses its skill once it reaches the value NRMSE $=0$, at which point the RMSE in the forecasted time series is equal to the standard deviation of the truth. The pattern correlation loses its skill when Corr$<0.5$.

\subsection{Forecasts Using Different Inputs}

\begin{figure}
\centering
\includegraphics[width=15.5cm]{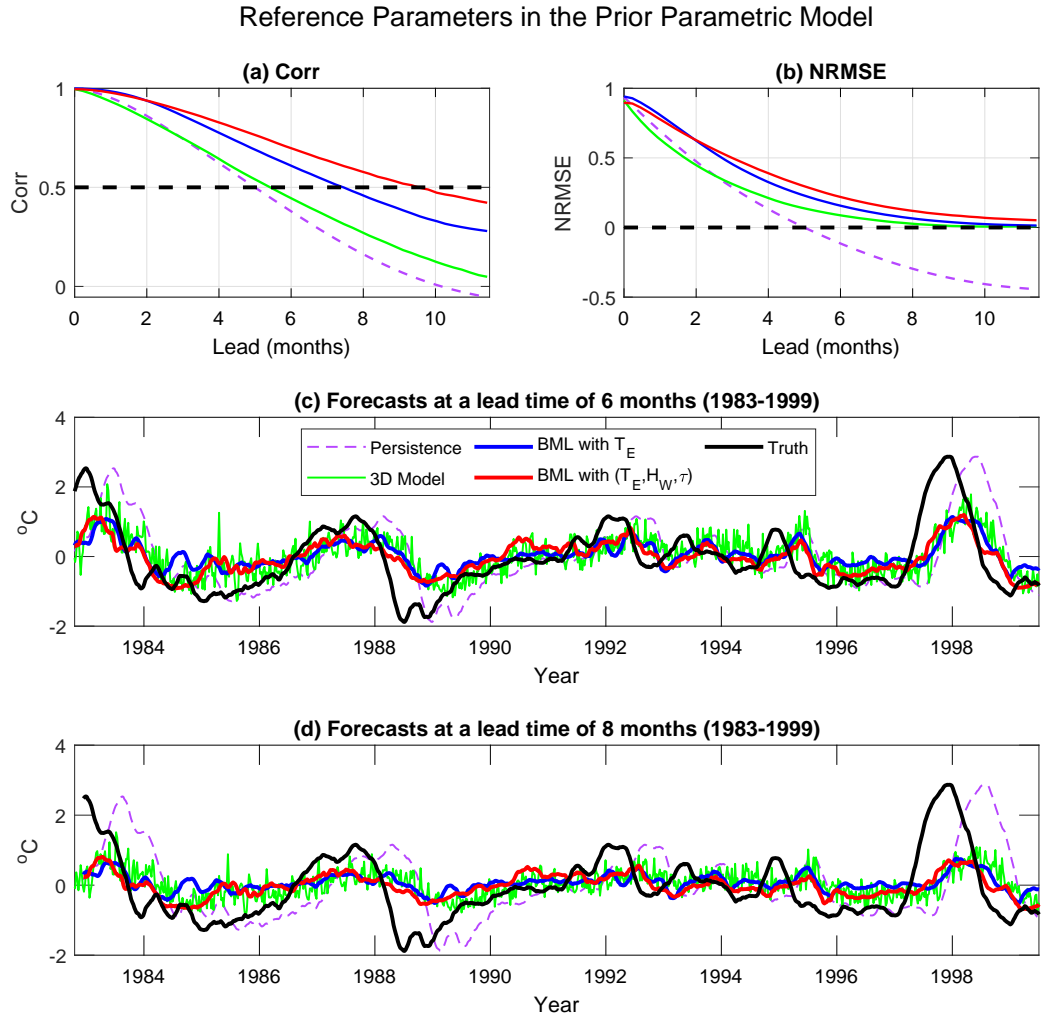}
\caption{Comparison of the forecasts of different inputs. Panels (a)--(b): skill scores of the persistence forecast (purple), the ensemble forecast using the 3D parametric model (green), the BML forecast using only $T_E$ as the input (blue), and the BML forecast using $(T_E, H_W, \tau)$ as the input (red). Panels (c)--(d): Comparison of the truth (black) and different forecasts at lead times of $6$ and $8$ months, respectively, during the 1982-1999 period. The predicted time series during the 2001-2020 period is included Figure \ref{standardparampriorEScovariates_SI_2020}.}\label{standardparampriorEScovariates}
\end{figure}

Figure \ref{standardparampriorEScovariates} shows the forecast skill of different inputs, using the BML algorithm, and compares them to the classical  approach of training the network for a fixed number of epochs.

The skill scores in Panels (a)--(b) indicate that the persistence forecast (purple curves) is skillful only up to $5$ months. The traditional ensemble forecast (green curves) based on the 3D model \eqref{priormodel} outperforms the persistence forecast and it remains effective for up to about $6$ months. Yet, this purely model-based ensemble forecast is far less skillful than the BML forecast. Even in the situation with only $T_E$ being the input (blue curves), the BML forecast remains skillful for up to nearly $7.5$ months. This indicates the advantage of the machine learning forecast over the traditional ensemble forecast based on parametric models.

Next, if all the three variables $(T_E, H_W, \tau)$ are used as the input, then the BML forecast skill can be further extended from $7.5$ to $9.5$ months (red curves). The WWBs are known to be important triggering effects of the El Ni\~no \cite{seiki2007westerly, lian2014effects, hu2014impact, thual2016simple} and this is reflected in the BML forecast shown here. Specifically, Panels (c)--(d) of Figure \ref{standardparampriorEScovariates} indicate that the forecast with the additional input of the wind bursts can capture the timing of the strong El Ni\~no (1987-1988 and 1997-1998) more accurately. It is important to note that since the wind bursts lie in the intraseasonal time scale, only the wind burst data during the past $8$ days is used as the BML input, which is much shorter than the $8$-week input data of $T_E$ and $H_W$. Such a multiscale feature for the input is crucial in facilitating the effective BML forecast.

We also found that the BML forecast with $(T_E, H_W)$ as input has about the same skill as the forecast where $T_E$ is the only input (see Figure \ref{standardparampriorEScovariates_SI_HW} in the \emph{Supporting Information}). A plausible explanation for this result is as follows. According to the celebrated recharge-discharge theory \cite{jin1997equatorial}, the variables $T_E$ and $H_W$ are the two components of a coupled system that forms an oscillator. Since the input of the BML contains the historic data up to the past $8$ weeks, the information in $H_W$ has been characterized by such historic data of $T_E$ according to the Takens' delayed embedding theorem \cite{takens1981detecting}. Thus, including the additional variable $H_W$ as the input provides little improvement for the BML forecast. Note that the use of both the current and the past data as the BML input shares a similar mechanism as the delayed oscillator theory \cite{suarez1988delayed, battisti1989interannual}, which involves only a single variable but contains the historical values in the differential equations.

\subsection{Reducing the Spring Predictability Barrier}
\begin{figure}
\centering
\includegraphics[width=14.5cm]{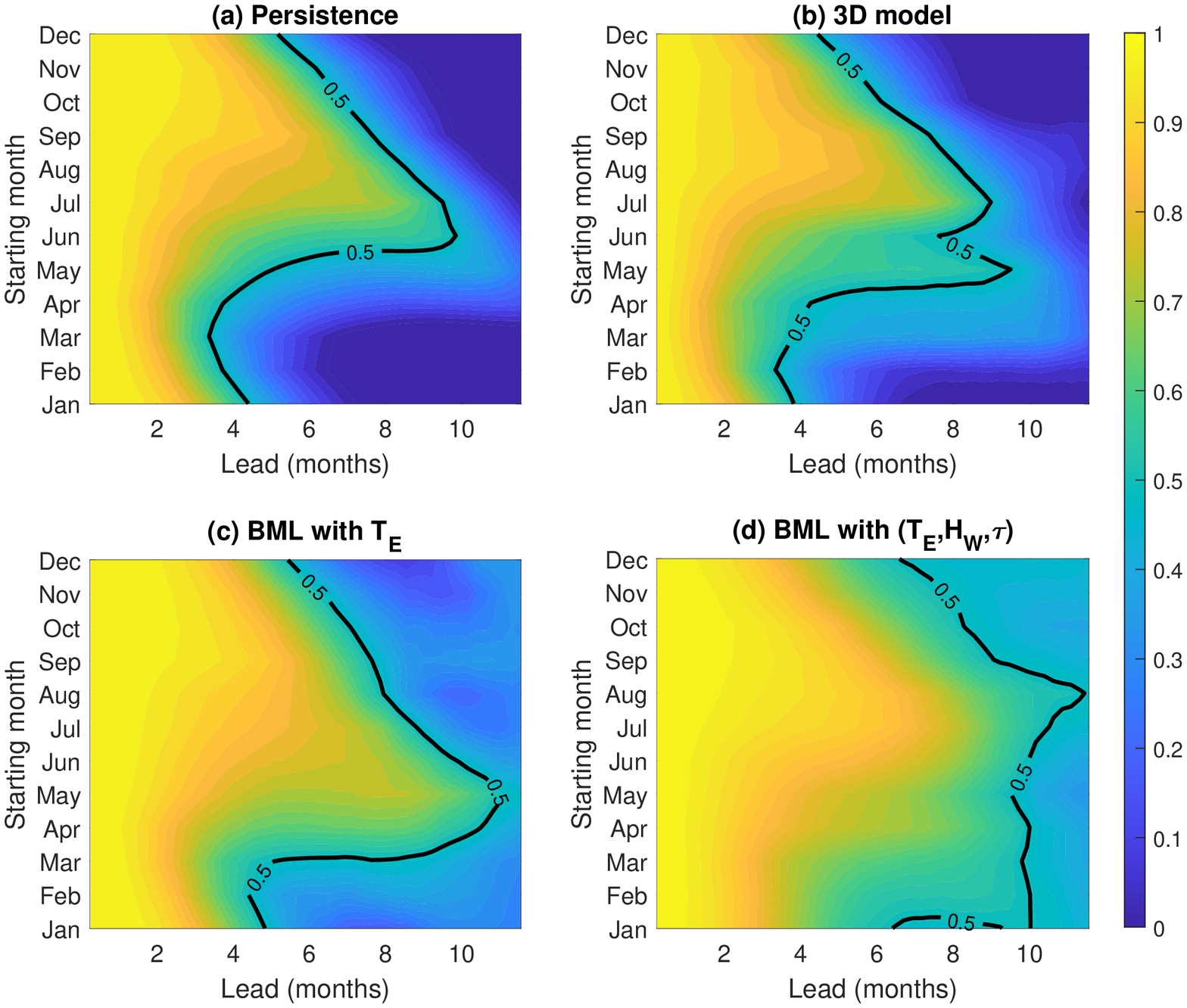}
\caption{Seasonal prediction: pattern correlation as a function of starting month (y-axis) and lead time (x-axis). Panel (a): persistence. Panel (b): ensemble forecast using the 3D parametric model. Panel (c): the Bayesian machine learning forecast using only $T_E$ as the input. Panel (d): Bayesian machine learning forecast using $(T_E, H_W, \tau)$ as the input.  }\label{Seasonal_Prediction}
\end{figure}

Figure \ref{Seasonal_Prediction} shows the forecasts starting from different months. Both the persistence and the ensemble forecast using the 3D model suffer from the so-called spring predictability barrier \cite{duan2013spring, barnston2012skill}. In particular, the forecast is quite challenging when starting from February, March, or April, where the skillful forecast only lasts for up to $4$ months.  The BML forecast with a single variable $T_E$ as the input slightly improves the forecast skill starting from February and March while the useful forecast starting from April is significantly extended to $9$ months. The more improved forecast results are provided by the BML with $(T_E, H_W, \tau)$ being the input. In this case, the skillful forecast lasts for nearly $10$ months starting from any time between February to August. In particular, the Corr remains above $0.75$ at a lead time of $4$ months when the starting time is boreal spring and the threshold of Corr $=0.5$ is nearly uniform for different starting months. Since the wind accounts for a large portion of the uncertainty in the ENSO forecast, the improved results here are consistent with the viewpoints in \cite{yu2009dynamics, mu2007kind} that exploited an advanced approach to study different initial uncertainties in affecting the ENSO forecast. Notably, the forecast skill starting from April is significantly improved using the BML. This finding supports the argument in \cite{kirtman2001current, duan2013spring} that indicates the forecast starting in this season is relatively easy and there is no notable spring predictability barrier phenomenon.

\subsection{Comparison to Standard Machine Learning training Procedures}\label{Sec:Results_ComparisonML}
\begin{figure}
\centering
\includegraphics[width=15.5cm]{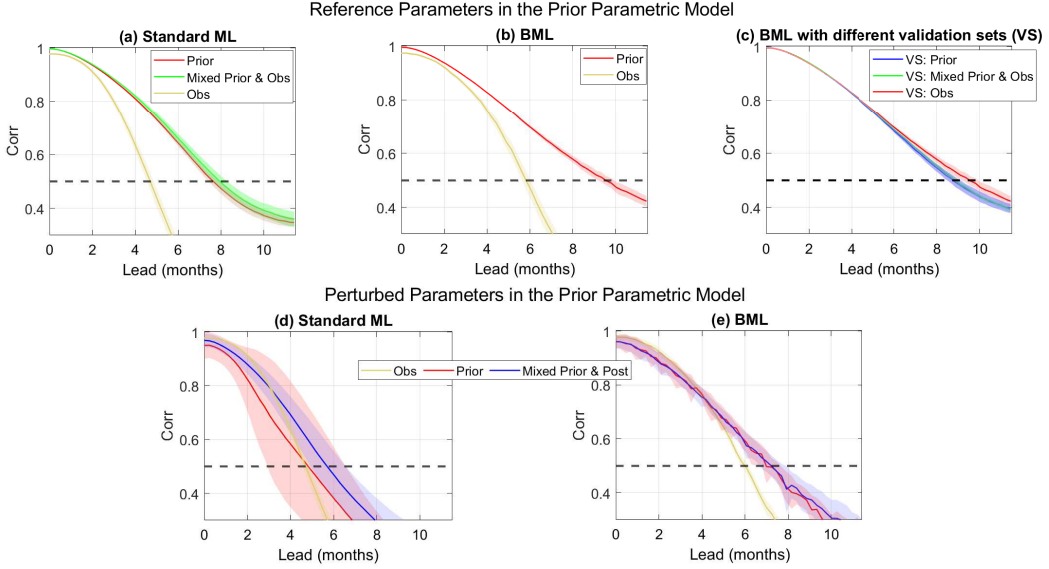}
\caption{Panels (a)--(c): Comparison of different setups in the machine learning (ML) algorithms. Panel (a): forecasting using the standard ML algorithm. Here, no validation criterion (or step 2 of BML)  is used to guide the training in \eqref{Training_NN_D}. The red and yellow curves indicate the forecasts, in which only the prior data and only the short observations are used for training, respectively. The green curve shows the forecast skill, where the training data is the concatenation of the long prior data and the short observations. Panel (b): forecasting using the BML algorithm. The red curve is the one that is the same as that in Figure \ref{standardparampriorEScovariates}. It exploits the prior data for the parameter updating \eqref{Training_NN_D} in the training process while the short observations are adopted for the data-driven validation \eqref{Training_NN_O} as the stopping criterion. The yellow curve uses the short observations for both parameter updating and the data-driven validation. Panel (c): forecasting using the BML algorithm but using different validation sets. The red curve is the same as that in Panel (b). The blue one uses the prior data for validation \eqref{Training_NN_O} while the green one uses the mixed prior data and the observations for validation. The shading area shows the uncertainty (95\% confidence interval) based on $10$ repeated experiments, which contain different realizations of the prior data.
Panels (d)--(e): Comparison of the standard and the BML algorithms in the situation with random perturbed parameters in the prior parametric model. The red and yellow curves indicate the forecasts, in which only the prior data and only the short observations are used for training, respectively. The blue curve shows the forecast, where the training data is the concatenation of the prior and posterior time series. The shading area shows the uncertainty based on $10$ repeated experiments, which contain different realizations of the prior data.
In all the panels, the input variables are
 $(T_E,H_W,\tau)$. }\label{Methods_Comparison_All}
\end{figure}

Now we report more numerical experiments to get a better sense of the BML algorithm. All the experiments in this subsection employ the variables $(T_E,H_W,\tau)$ as the input. Note that the SGD can approach a local minimum of the loss function, which is generally nonconvex, so the initial value of $\boldsymbol\theta$ is a source of uncertainty in the neural network training. Another source of the uncertainty comes from the random realizations of the prior time series. To account for these uncertainties, we run the training and forecasting algorithm $10$ times for each experiment with different initial guesses of $\boldsymbol\theta$ and different realizations of the prior time series. The shading area in Figure \ref{Methods_Comparison_All} shows the 95\% confidence interval of the forecast skill for each experiment.

Figure \ref{Methods_Comparison_All} compares the BML forecast (panel (b)) with the forecast of the same neural network architecture trained in a standard fashion (panel (a)), using  \eqref{Training_NN_D} for a fixed number of epochs instead of the validation Step 2. The red curve in Panel (a) shows that the neural network trained for a fixed number of epochs on only the prior time series results in a forecast that remains skillful for only $7.5$ month, which is much shorter than the BML forecast ($9.5$ months) shown in Panel (b). Importantly, as is shown by the green curve in Panel (a), in the absence of the data-driven stopping criterion \eqref{Training_NN_O}, even if the observational and the prior data are both included in \eqref{Training_NN_D} for training the network, the resulting forecast skill has little improvement. These findings indicate the importance of using the limited observational data to validate the proposals using \eqref{Training_NN_O}. On the other hand, as anticipated, when the prior time series is replaced by the short observations in \eqref{Training_NN_D} for updating $\boldsymbol\theta$, the small amount of training data leads to an extremely unskillful forecast (the yellow curves).

One important issue is to understand the robustness of the proposed scheme to the choice of the validation set in applying the second step of BML. In Panel (c) of Figure \ref{Methods_Comparison_All}, it is shown that if the short observations are replaced by a prior time series from a time interval independent from the one used in \eqref{Training_NN_D}, then the skillful forecast becomes less than $8.5$ months (blue curve). Notably, the confidence interval associated with such an experiment has no overlap with that using the short observations as the validation set (red curve) at the Corr $=0.5$ threshold, which indicates the statistical significance of the difference between these two methods. Another test is to first mix the long prior data and the short observational data, which is then followed by splitting the mixed data into the training and validation sets for Eq.~\eqref{Training_NN_D} and Eq.~\eqref{Training_NN_O}, respectively. However, since the prior data is much longer than the observations, such an approach (green curve) leads to essentially the same forecast result as the one without the observational data (blue curve). These facts highlight the importance of the data-driven validation \eqref{Training_NN_O} that serves as the likelihood function in the BML framework.

\subsection{Sensitivity of BML Algorithm Under Perturbations of the Prior Model}\label{Sec:Results_ModelError}
The prior time series in all the tests that we have shown so far were generated from the 3D parametric model \eqref{priormodel} with the reference parameters \eqref{ParameterValues}. While the modeling error prohibits this model to reflect more complex features, such as the spatiotemporal patterns of the ENSO diversity, it qualitatively reproduces the characteristics of the Ni\~no 3 SST time index (as shown in Fig.~\ref{ENSO_Comparison}). It is therefore critical to understand the sensitivity of the BML algorithm to the choice of prior models. With this goal in mind, we randomly perturb the reference parameters and check the forecast skills of the standard machine learning and the BML algorithms, when both models are trained using the training data from the perturbed prior models. We repeat the experiments with the perturbed parameters $10$ times. In each experiment, we add random Gaussian noise to each of the reference parameters in \eqref{ParameterValues}, where the random Gaussian noise has zero mean and variance $0.8$. The model simulation with the perturbed parameters is illustrated in the \emph{Supporting Information}.

In Panels (d)--(e) of Figure \ref{Methods_Comparison_All}, we show the averaged skill score over the $10$ outcomes for each experiment as well as the associated uncertainty (shading area), which mimic the scenario of the multi-model forecast except that each ``model'' here is a machine learning forecast, obtained based on training data generated by the prior model with different parameters.  Panel (d) shows the results using the standard machine learning forecast algorithm (red curve). Due to the relatively large model error in the prior model, the averaged skill score using the prior time series as the training data is $5$ months, which is only slightly better than the forecast using the short observational data for training (yellow curve). The associated uncertainty (red shading area) using the standard machine learning forecast is also quite large.
The forecast results using the standard machine learning algorithm can be improved if the prior time series are concatenated with the so-called posterior time series. These posterior time series are obtained by using a recently developed sampling algorithm \cite{chen2020can}, which exploits data assimilation to sample a collection of time series based on the prior model and the short observations. See the \emph{Supporting Information} for technique details. Using these posterior time series as the training data, the skillful forecast of the standard machine learning algorithm can be extended to about $6$ months together with a reduction of the forecast uncertainty.

In contrast, despite the large model error in the prior model with the perturbed parameters, the forecast skill score using the BML, as is shown in the red curve in Panel (e) of Figure \ref{Methods_Comparison_All}, is still skillful up to $7$ months. Also, the uncertainty (red shading area) is significantly smaller than the uncertainty of the standard machine learning forecast, which indicates that the BML algorithm is robust. These merits are due to the data-driven validation \eqref{Training_NN_O}, which significantly reduces the model error in the prior time series. One interesting finding is that even if one includes posterior time series obtained using a class of data assimilation schemes (See the \emph{Supporting Information} for details) into the training data for updating $\boldsymbol\theta$ of \eqref{Training_NN_D} (blue curve in Panel (e)), the skill score remains almost the same as that using only the prior data in \eqref{Training_NN_D}. This is because the BML algorithm has already taken into account the combined information from the prior model and the observations through the two steps in the training procedure, which captures the information contained in the posterior time series with less uncertainty. This suggests another practical advantage of BML: it does not depend on posterior time series produced by a separate ``data assimilation'' scheme that may involve non-negligible computational efforts, depending on the complexity of the prior model.

\section{Conclusion}\label{Sec:Conclusion}
In this article, a simple and efficient BML training algorithm, which exploits only short observational time series and an approximate parametric model, is developed to provide effective predictions of the Ni\~no 3 SST index. The BML forecast significantly outperforms the model-based ensemble predictions and standard machine learning forecasts. It also overcomes the spring predictability barrier to a large extent. The BML algorithm allows a multiscale input consisting of both the SST and the wind bursts, which improved prediction skills over the forecasts where SST is the sole variable. Finally, the BML algorithm also reduces the forecast uncertainty and produces robust results even when the training data is generated by prior models that are significantly perturbed from its reference parameter.

The primary goal here is to provide a new and efficient training framework that facilitates a predictive machine learning model even when  there is a small amount of available training data. The main idea is to supplement the training strategy with a set of computer-generated data from a model to reduce the large variance error usually present in an ML model trained using a standard training procedure. While the BML may not be better than the standard training procedure when a large dataset becomes available, it is worthwhile to try and determine the length of  observational data required so that the two methods are comparable. This is a complicated study to undertake since it may depend on the class of neural network models and the choice of prior models. Unfortunately, in the context of SST prediction, such a study may not be conclusive due to the shortage of available data and unknown underlying dynamics. We plan to address this issue in a more controlled environment, where the underlying dynamics are known. Finally, while the framework is rather simple and can be used in any application where there are a small number of observations, the effectiveness of the method depends on the choice of prior model. In real applications, since the underlying dynamical model is not available, the main task is to identify an effective prior model. In the context of SST prediction with time series training data, we have shown the effectiveness of BML over standard training procedure using the SDE in \eqref{priormodel} as prior.
For future work, we plan to inspect whether other variables, such as the tropical and extratropical precursors \cite{chen2020enhancing, boschat2013extratropical} can improve the SST forecast. It is also interesting to use operational models as the prior to test the BML algorithm. Another natural future work using the new Bayesian machine learning framework is to skillfully forecast spatiotemporal patterns of the ENSO diversity.

\bibliography{references.bib}

\newpage
\section*{\LARGE Supporting Information for ``A Bayesian Machine Learning Algorithm for Predicting ENSO Using Short Observational Time Series''}
\setcounter{section}{0}
\noindent\textbf{Contents of this file}
%%%Remove or add items as needed%%%
\begin{enumerate}
\item Comparison of the Simulation from the 3D Model with the Observations
\item More Results of the BML Forecasts with the reference Parameters in the Prior Model
\begin{enumerate}
  \item BML Forecasts Using Different Inputs
  \item Forecasts Using Different Inputs 
\end{enumerate}
\item Details of the BML Forecasts with Perturbed Parameters in the Prior Model
\begin{enumerate}
  \item The Model with Perturbed Parameters
  \item Discussion of the Conditional Sampling Method that Generates the Posterior Time Series
\end{enumerate}
%if Tables are larger than 1 page, upload as separate excel file
\end{enumerate}
\noindent\textbf{Additional Supporting Information (Files uploaded separately)}
\begin{enumerate}
\item Source code for the BML training and forecasts.
\end{enumerate}

\section{Comparison of the Simulation from the 3D Model with the Observations}
\begin{figure}[h]
\hspace*{-0.8cm}\includegraphics[width=19.0cm]{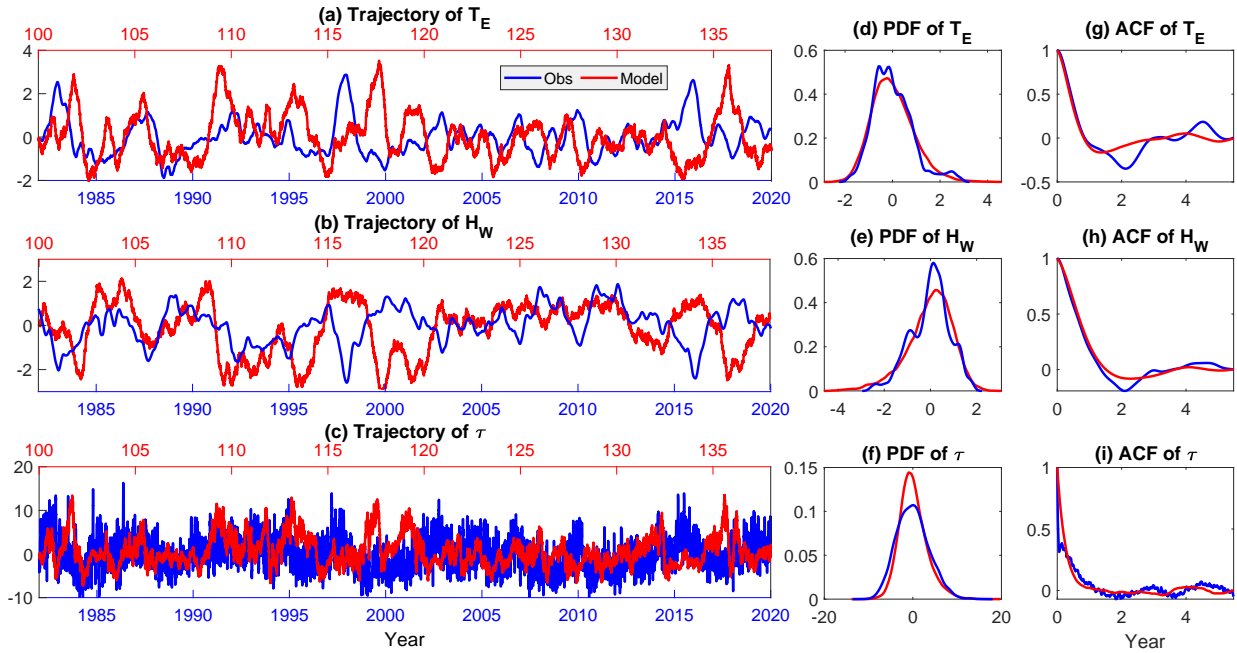}
\caption{Comparison of the observations (blue)  with the simulation from the three-dimensional model \eqref{priormodel} with the reference parameters \eqref{ParameterValues} (red). Panels (a)--(c): time series trajectories. Panels (d)--(f): PDFs. Panels (g)--(i): ACFs. Note that Panels (a)--(c) simply show one random realization of the model and thus there is no one-to-one correspondence between the blue and red trajectories (model v.s. observations). The x-axis on the top of each panel is for the model simulation while the one on the bottom is for the observations.}\label{ENSO_Comparison}
\end{figure}

Panels (a)--(c) of Figure \ref{ENSO_Comparison} show a comparison between the model trajectories and the observational time series. Here the reference parameters \eqref{ParameterValues} are used in the model \eqref{priormodel}. The results here indicate qualitatively consistent path-wise behavior. Panels (d)--(i) compare the PDFs and the ACFs, and show that the observed non-Gaussian statistics of all of the three variables are reproduced by the model to a large extent. As seen Figures~\ref{standardparampriorEScovariates} and \ref{standardparampriorEScovariates_SI_2020}, this model can reproduce the climatological statistics and trajectories that qualitatively reflect the time scale of the truth; but, its predictive skill is less effective than the proposed BML algorithm due to the model error.

\section{More Results of the BML Forecasts with the Reference Parameters in the Prior Model}
\subsection{BML Forecasts Using Different Inputs}
\begin{figure}
%\hspace*{-0.5cm}
\centering
\includegraphics[width=15.5cm]{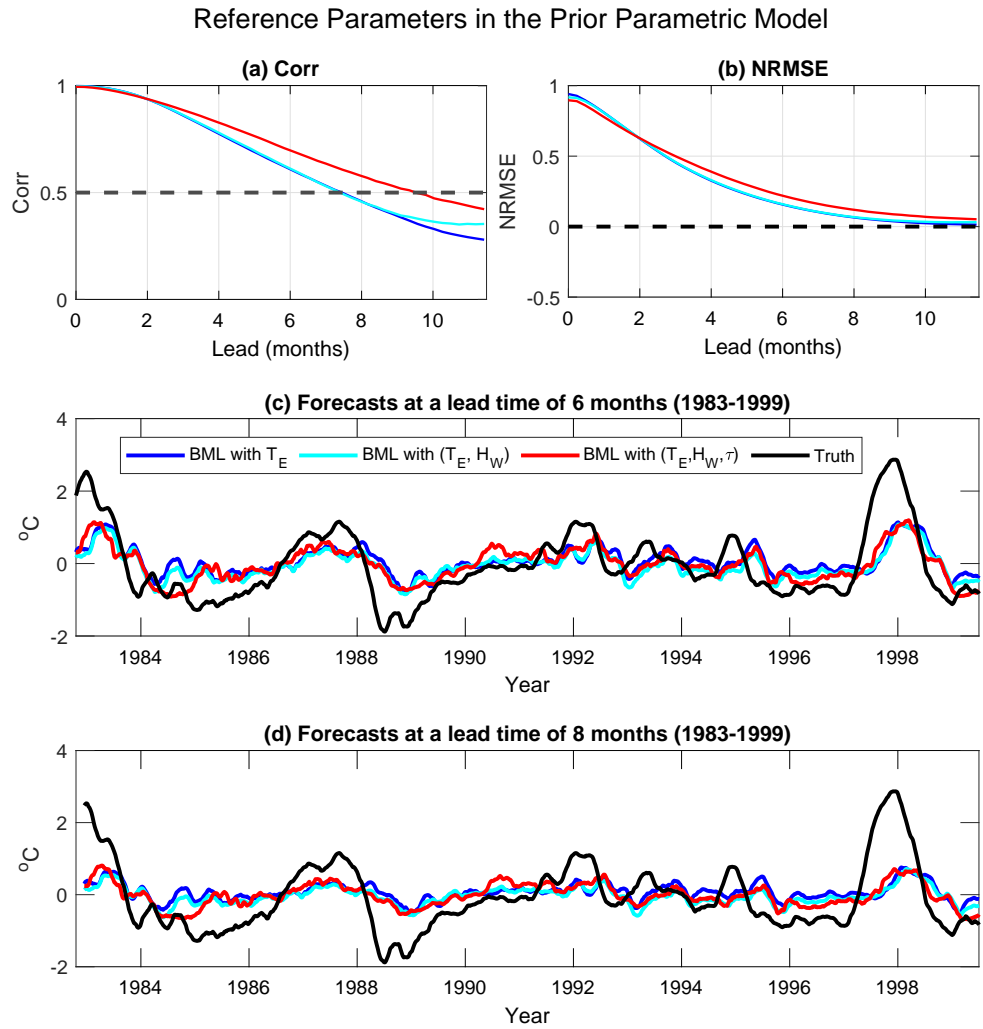}
\caption{Similar to Figure \ref{standardparampriorEScovariates} in the main text, except that the BML forecast using $(T_E, H_W)$ as the input (cyan) is included. To clarify the presentation, the persistence and 3D model ensemble forecast have been ignored here.}\label{standardparampriorEScovariates_SI_HW}
\end{figure}

Figure \ref{standardparampriorEScovariates_SI_HW} shows   the BML forecast using different inputs. It is similar to Figure \ref{standardparampriorEScovariates} in the main text, except that the BML forecast using $(T_E, H_W)$ as the input (cyan) is also included. The results here show that the forecast with input $(T_E, H_W)$ is similar to the one with $T_E$ as the only input.

\subsection{Forecasts Using Different Inputs}
\begin{figure}
%\hspace*{-0.5cm}
\centering
\includegraphics[width=15.5cm]{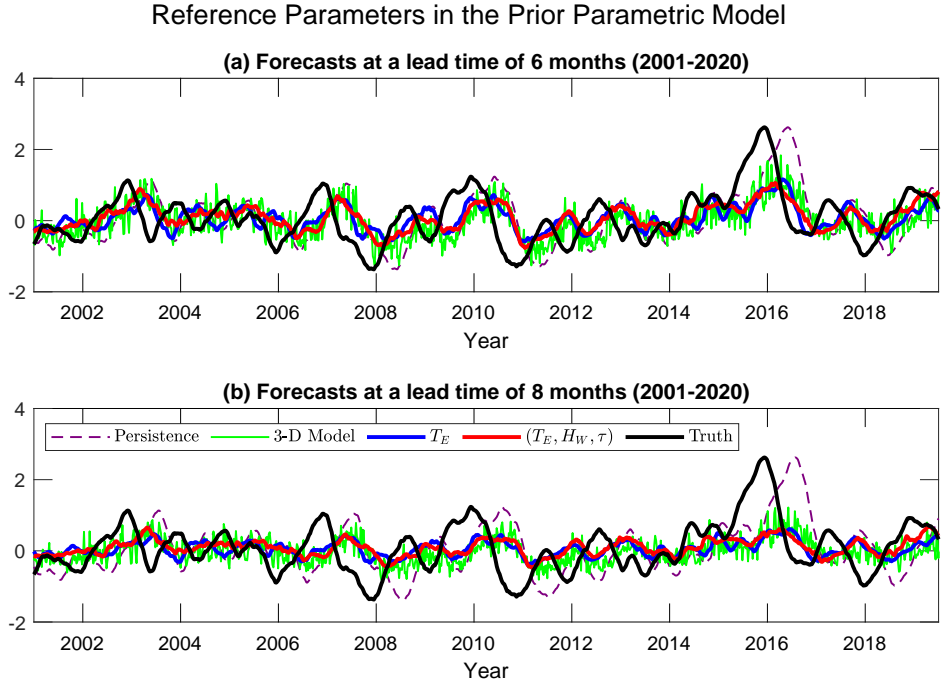}
\caption{Similar to Panels (c)--(d) of Figure \ref{standardparampriorEScovariates} in the main text, except the results are in the 2001-2020 time period.}\label{standardparampriorEScovariates_SI_2020}
\end{figure}

Figure \ref{standardparampriorEScovariates_SI_2020}, which is similar to Panels (c)--(d) of Figure \ref{standardparampriorEScovariates} in the main text, shows the forecasts obtained from different inputs. 
The only difference is that the results in Figure \ref{standardparampriorEScovariates_SI_2020} are in the 2001-2020 period.

\section{Details of the BML Forecasts with Perturbed Parameters in the Prior Model}

\subsection{The Model with Perturbed Parameters}
We still adopt the same model as in \eqref{priormodel} when considering the model error case, But we add independent Gaussian random noise with mean $0$ and variance $.8$ to each parameter. Figure \ref{TE_nonstandard_series_statistics} show $5$ perturbed models with different random parameters (red curves) along with the observations (blue curves). For simplicity, the random number seeds are fixed here. Therefore, the main difference in these perturbed parameter cases is the amplitude, phase and noise levels in the time series. It is clear that some imperfect models (e.g., Cases \#2 and \#3) are qualitatively reasonable while some other perturbed models (e.g., Cases \#4 and \#5) are quite different from the observations. This situation mimics the multi-model forecast scenario, where the differences between the models can be quite dramatic.

\subsection{Discussion of the Conditional Sampling Method that Generates the Posterior Time Series}
In this section, we briefly discuss the conditional sampling method that generates the posterior time series that are used in assessing the robustness of BML (see blue curves in panels (c)-(d) in Figure \ref{Methods_Comparison_All}. In particular, we will sample the following conditional distribution,
\begin{equation}%\eqnum{S1}
p(\mathbf{v}(t)|\mathbf{u}([0,T])), \quad 0\leq t\leq T,\notag
\end{equation}
where $\mathbf{v}(t)$ and $\mathbf{u}([0,T])$ are the time series of the prior model and the observations,respectively on the training time interval $[0,T]$. 
Numerically, we sample the above conditional distribution by solving a system of stochastic differential equations resulting from a nonlinear version of the Kalman smoother applied to the prior model in \eqref{priormodel}. The mathematical details are shown in \citep{chen2020can}. Therefore, the sampled trajectories have the same length as the observations and they lie in the same time interval. In a nutshell, we effectively leverage the randomness in the prior model \eqref{priormodel}, reflected by the stochastic noises, to generate multiple trajectories that contain the information from both the observations and the prior model.

%\jh{Comments on the paragraph above: Should Eqn (1) be replaced with the smoothing version? I also reshuffled the discussion and hope it clarifies the scheme without presenting more detail. There is also a conflict of labelling. The bayesian eqn here was originally labelled exactly as (1) which is for prior; so I relabelled it as S1 with eqnum.}\nc{[I would prefer not to show the details of the smoothing version in (1). The current equation (1) should be good enough for the geophysical people. We anyway said in the text we use a smoother.]}

In the model with reference parameters, the sampled trajectories highly resemble the observations. See Panels (a)--(b) in Figure \ref{Sampled_Paths}.

\begin{figure}
\hspace*{-0.7cm}
%\centering
\includegraphics[width=19cm]{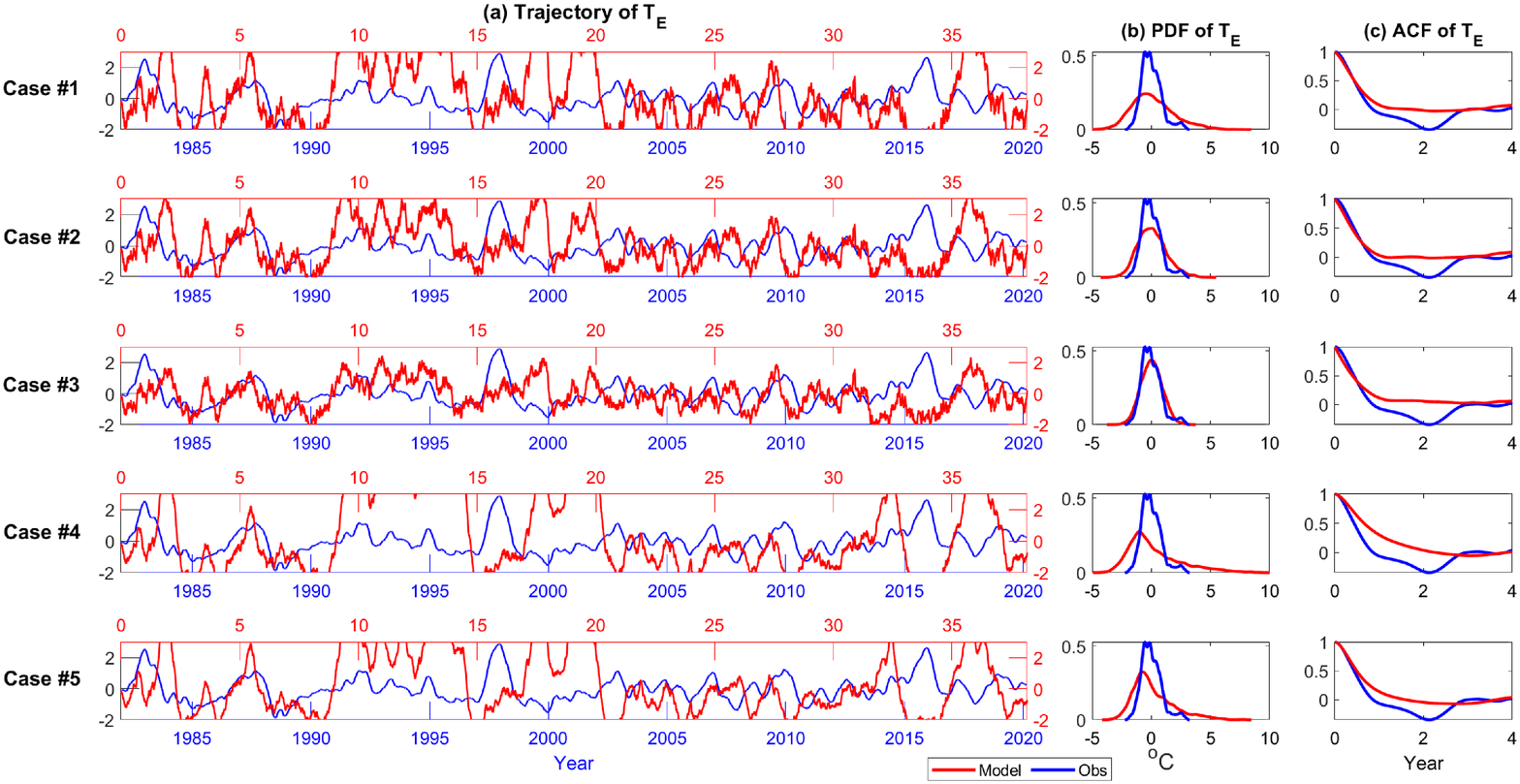}
\caption{The model \eqref{priormodel} with perturbed parameters. The five rows show five groups of perturbed parameters. The three columns compare the time series, PDFs and ACFs with the truth. }\label{TE_nonstandard_series_statistics}
\end{figure}

\begin{figure}
\hspace*{-1cm}\includegraphics[width=19cm]{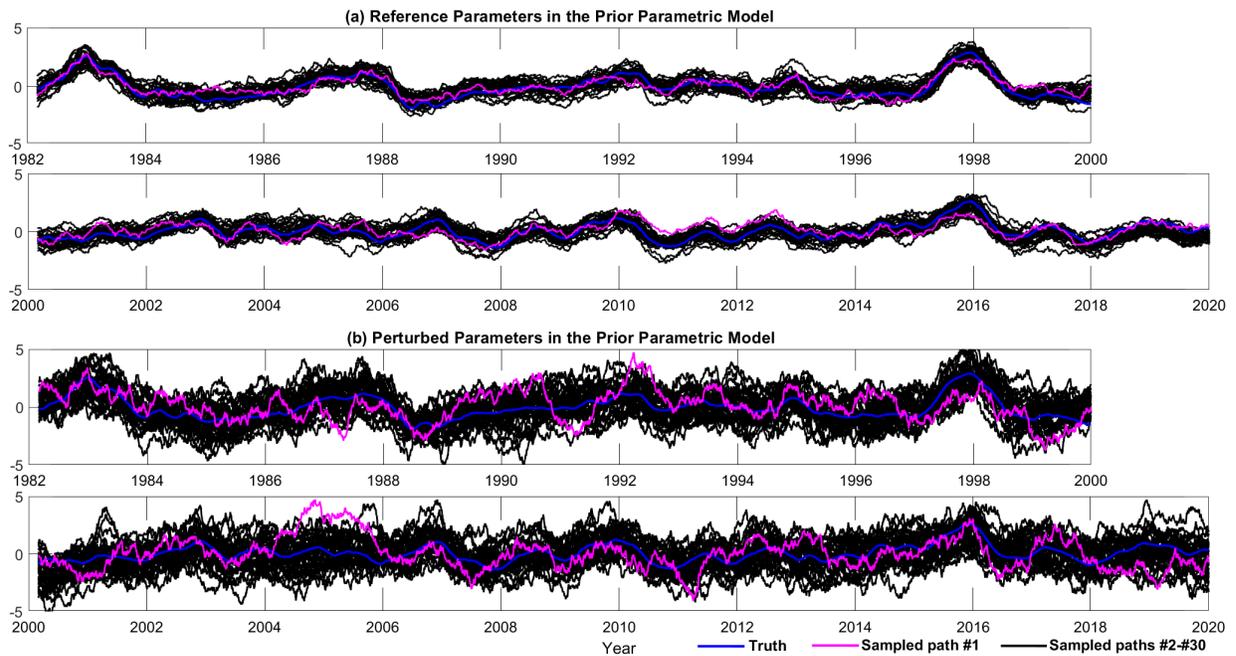}
\caption{Comparison of the observed SST (blue) and the 30 sampled of posterior time series (magenta and black). One sampled trajectory is marked by magenta color for the convenience of comparing with the observation. Panel (a): sampled trajectories using the
smoothing equation corresponding to the prior model \eqref{priormodel} equipped with the reference parameter.
Panel (b): Same as (a) except that the prior model is perturbed. Here, we show sampled trajectories corresponding to Case \#1 in Figure \ref{TE_nonstandard_series_statistics}.
% \jh{Comment: I just try to make sure the reader understand this picture corresponds to the posterior time series, which was not described in previous caption; as the title of the panels do not reflect this.}
}\label{Sampled_Paths}
\end{figure}

In the situation with perturbed parameters, we focus on Case \#1 shown in Figure \ref{TE_nonstandard_series_statistics}. The perturbed parameters are $\alpha_1=2.2603$, $\alpha_2=.8676$, $d_\tau=4.1833$, $d_T=d_H= 2.1393$, $\omega_u=-1.3890$, $\sigma_T=\sigma_H=2.1324$. Furthermore, the original $\sigma_\tau$ parameter is multiplied by $0.6715$. As shown in the first row of Figure \ref{TE_nonstandard_series_statistics} (red color), the prior time series generated from the perturbed model exhibits It can be seen from panel (b) of Figure \ref{Sampled_Paths}, that the sampled trajectories have less error than the prior time series. This is the fundamental reason that the traditional neural network forecast skill can be improved when the prior and posterior time series are used together for training (Panel (d) of Figure \ref{Methods_Comparison_All}). Note that the length of the posterior time series is the same as the length of the short time observations. Therefore, although the multiple posterior time series plays an important role in reducing the bias (model error) in the training data when compared with the training based only on the prior time series, these short posterior time series may not cover the entire solution space associated with the true underlying dynamics. The prior time series can be included in the training dataset to expand the solution space, thereby compensating for this shortcoming of the posterior time series.  

It is worthwhile to point out that in the simulation above, since the prior model in \eqref{priormodel} has a special structure (it is conditionally Gaussian), the resulting smoothing equation can be realized analytically \citep{chen2020can}. Unfortunately, such structure is not amenable to generalization, and applying the conditional sampling algorithm for general nonlinear systems in high dimensional space requires additional computational efforts that can be expensive.
For example, when ensemble Kalman smoother is used, the number of the ensemble members must be increased (exponentially) in the Bayesian update step to retain accurate numerical solutions. Our numerical results (compare the blue and red curves in panel (e) of Figure.~\ref{Methods_Comparison_All}) suggest that the proposed BML algorithm can readily account for the information in the posterior time series without having to realize it with an additional smoother algorithm, and thus, additional computational cost can be avoided.

\clearpage

\end{document}